\documentclass[11pt,a4paper]{article}
\usepackage{jcappub}

\usepackage{pdflscape}
\usepackage{amsmath}
\usepackage{amssymb}
\usepackage{dcolumn}
\usepackage{bm}
\usepackage{color}
\usepackage{epsfig}
\usepackage{amsfonts}
\usepackage{graphicx}
\usepackage{subfigure}
\usepackage{dcolumn}

\newcommand{\be}{\begin{equation}}
\newcommand{\ee}{\end{equation}}
\newcommand{\bea}{\begin{eqnarray}}
\newcommand{\eea}{\end{eqnarray}}

\renewcommand{\[}{\left[}

\def\be{\begin{equation}}
\def\ee{\end{equation}}
\def\bea{\begin{eqnarray}}
\def\eea{\end{eqnarray}}

\begin{document}

\title{Higher Order Lagrangians inspired by the Pais-Uhlenbeck Oscillator and their cosmological applications}

\author[a]{Gustavo Pulgar}

\author[a]{Joel Saavedra}

\author[a]{Genly Leon}

\author[b]{Yoelsy Leyva}

\affiliation[a]{Instituto de F\'{\i}sica, Pontificia Universidad  Cat\'olica
de Valpara\'{\i}so, Casilla 4950, Valpara\'{\i}so, Chile}
\affiliation[b]{Departamento de F\'{\i}sica, Facultad de Ciencias, Universidad de Tarapac\'a, Casilla 7-D, Arica, Chile}

\emailAdd{gustavopulgar@gmail.com}

\emailAdd{joel.saavedra@ucv.cl}

\emailAdd{genly.leon@ucv.cl}

\emailAdd{yoelsy.leyva@uta.cl}

\date{\today}
\abstract{We study higher derivative terms associated with scalar field cosmology. We consider  a coupling between the scalar field and the geometry inspired by the Pais-Uhlenbeck oscillator, given by $\alpha\partial_{\mu}\partial^{\mu}\phi\partial_{\nu}\partial^{\nu}\phi.$  We investigate the cosmological dynamics in a phase space. For $\alpha>0$, we provide conditions for the stability of de Sitter solutions. In this case the crossing of the phantom divide  $w_{DE}=-1$ occurs once; thereafter, the equation of state parameter remains under this line, asymptotically reaching towards the de Sitter solution from below. For $\alpha<0,$ which is the portion of the parameter space where in addition to  crossing  the phantom divide,  cyclic behavior is possible, we present regions in the parameter space where, according to Smilga's classification the ghost  has benign or malicious behavior.}


\maketitle


\section{Introduction}
Scalar fields have been widely used in cosmology as good candidates to describe the early (inflationary era) and late universe (dark energy). When using a scalar field as the matter content of the
universe, the coupling between the scalar field and the geometry must be specified. In this sense, there is enough literature about the minimal  and  non-minimal coupling cases \cite{Linde:1983gd,Copeland:2006wr,Amendola:1999qq,Faraoni:2000wk,Ratra:1987rm,Liddle:1998xm,Halliwell:1986ja,Uzan:1999ch,Faraoni:2012bf,Bertolami:1999dp,Faraoni:1996rf,Boisseau:2000pr}. 
The standard action to describe a minimally coupled scalar field is given by
\begin{equation}
S=\int d{}^4x\sqrt{-g}\left(\frac{R}{2}-\frac{1}{2} g^{\mu \nu}\partial_{\mu}\phi\partial_{\nu}\phi-V(\phi)\right),\label{action1}
\end{equation}
where the scalar field is minimally coupled to gravity \footnote{This manuscript has units where $M^2 \equiv \frac{1}{8\pi G}=1$.}. We can consider another non-minimal coupling by adding one term to (\ref{action1}), given by
\begin{equation}
\int d{}^4x\sqrt{-g}\zeta R \phi^2,\label{action2}
\end{equation}
where the scalar field is now  coupled to the geometry through the Ricci scalar. 
A generalization of  action \eqref{action2} was investigated in \cite{Skugoreva:2014gka} by considering potentials $V(\phi)=\phi^n$ and $V(\phi)=\phi^{n_1}+\phi^{n_2}$, and couplings $-\zeta B(\phi) R$ (where $\zeta$ is the coupling constant) and $B(\phi)=\phi^N.$ The global picture of the phase space was investigated  by means of compact variables.  For some intervals of the slopes of the potential and the coupling function it was possible to find some exact solutions. In reference \cite{Aref'eva:2012au}, a negative cosmological constant was added to \eqref{action2} . This allowed for a quasi-cyclic universe evolution with the Hubble parameter oscillating from positive to negative values.  Either one or several cycles can occur, depending on the initial conditions, before becoming negative forever. 
Some very close models are the so-called quinstant models (non-minimally coupled scalar field with the addition of a negative cosmological constant), which were discussed from a dynamical systems point of view in \cite{Leon:2009ce}. In addition, from both qualitative and observational viewpoints, other Dark Energy models, e.g., the quintom paradigm, were reviewed in \cite{Leon:2009ce} and new results were added to the state of art.  

To continue presenting alternatives, a scalar field can be coupled to the matter sector by adding a term to (\ref{action1})  of the form \cite{Kaloper:1997sh}
\begin{equation}
\int d{}^4x\sqrt{-g} \Omega(\phi)^{-2} \mathcal{L}(\chi, \nabla \chi, \Omega(\phi)^{-1}g_{\mu \nu}),\label{action2_1}
\end{equation}
where $\Omega(\phi)^{-2}$ is the coupling function, $\mathcal{L}$ is the matter Lagrangian, and $\chi$ is a
collective name for the matter degrees of freedom. The kinds of couplings been in \eqref{action2} and \eqref{action2_1} are related through conformal transformations (see \cite{Kaloper:1997sh} and references therein). 
In a recent paper \cite{Fadragas:2014mra}  a comprehensive review about theories based on the action \eqref{action2_1} was presented. 

It is well known that the more general scalar field Lagrangian with non-minimal coupling between the scalar field and the curvature and  which the same time produces second order motion equations, is the so-called Horndeski lagrangian \cite{Horndeski:1974wa}.  A special subclass, the Galileons, were constructed in \cite{Nicolis:2008in,Deffayet:2009wt,Deffayet:2009mn,Deffayet:2011gz}. In order for the field equations to satisfy the Galilean symmetry
$$\phi\rightarrow\phi+c, \;\partial_\mu \phi\rightarrow\partial_\mu \phi+b_\mu,\; c,b_\mu\, \text{constants},$$ in the Minkowskian limit, the four-dimensional
Lagrangian must be the sum of the Einstein-Hilbert lagrangian and four unique
terms consisting of scalar combinations of $\partial_\mu\phi$,
$\partial_\mu\partial_\nu\phi$ and $\Box\phi$, which are given by \cite{DeFelice:2011bh}: 
\begin{align}
&{\cal L}_{2} = K(\phi,X),\label{L2}\\
&{\cal L}_{3} = -G_{3}(\phi,X)\Box\phi,\\
&{\cal L}_{4} = G_{4}(\phi,X)\,
R+G_{4,X}\,[(\Box\phi)^{2}-(\nabla_{\mu}\nabla_{\nu}\phi)\,(\nabla^{\mu}
\nabla^{\nu}\phi)]\,,\\
&{\cal L}_{5} = G_{5}(\phi,X)\,
G_{\mu\nu}\,(\nabla^{\mu}\nabla^{\nu}\phi)\,\nonumber\\&\ \ \
\ \ \ \ -\frac{1}{6}\,
G_{5,X}\,[(\Box\phi)^{3}-3(\Box\phi)\,(\nabla_{\mu}\nabla_{\nu}\phi)\,
(\nabla^{\mu}\nabla^{\nu}\phi)+2(\nabla^{\mu}\nabla_{\alpha}\phi)\,(\nabla^
{\alpha}\nabla_{\beta}\phi)\,(\nabla^{\beta}\nabla_{\mu}\phi)]\,.\label{L5}
\end{align}
The functions $K$ and $G_{i}$ ($i=3,4,5$) depend on the scalar field $\phi$
and its kinetic energy
$X=-\partial^{\mu}\phi\partial_{\mu}\phi/2$, while
$R$ is the Ricci scalar, and $G_{\mu\nu}$ is the Einstein tensor.
$G_{i,X}$ and $G_{i,\phi}$ ($i=3,4,5$) respectively correspond to the
partial
derivatives of $G_{i}$ with respect to $X$ and $\phi$,
namely $G_{i,X}\equiv\partial G_{i}/\partial X$ and
$G_{i,\phi}\equiv\partial
G_{i}/\partial\phi$.

In \cite{Leon:2012mt}  the special case \begin{equation}
S=\int d^{4}x\sqrt{-g}\left[ \frac{R}{2}-\frac{1}{2}\partial^{\mu}\phi\partial_{\mu}\phi-V(\phi)
-\frac{1}{2}g(\phi)\partial^{\mu}\phi\partial_{\mu}\phi\Box\phi  +{\cal
L}_m\right]\,.\label{action_G}
\end{equation}
was investigated from the dynamical systems perspective. In this setup, we can find non-minimally coupled subclasses of Horndeski scalar-tensor theories that arises from the decoupling limit of massive gravity by covariantization \cite{deRham:2011by,Heisenberg:2014kea}. 

Now, in this paper, instead of investigating the Horndeski/Galileon class of models, we want to investigate a possible model that belongs to the more general theoretical form of the action i.e,  with more general coupling terms between the scalar field and the spacetime
curvatures, expressed as
{\small{
\begin{equation}
S=\int d{}^4x\sqrt{-g}\left\{f(R,R_{\mu \nu}R^{\mu \nu},R_{\mu \nu \lambda \rho}R^{\mu \nu \lambda \rho},\ldots)+ K(\phi,\partial_{\mu}\phi \partial^{\mu}\phi, \Box^2 \phi,R^{\mu \nu}\partial_{\mu}\phi\partial_{\nu}\phi,\ldots) -V(\phi)\right\},\label{action3}
\end{equation}}}
where $f$ and $K$ are arbitrary functions of the corresponding variables. Following the logic established above, the non-linear functions $f$ and $K$
provide more general non-minimal coupling between the scalar field and gravity. Of course these new coupling functions modify the usual Klein-Gordon
equation, and in contrast with the Horndeski/Galileon class, the field equation for the scalar field is no longer a second order differential equation. Some previous results in the literature are, for example: in Ref. \cite{Amendola:1993uh} where the authors used the coupling
$R^{\mu \nu}\partial_{\mu}\phi\partial_{\nu}\phi$, and found new analytical inflationary solutions;
in Ref. \cite{Capozziello:1999xt}, where the couplings $R\partial_{\mu}\phi\partial^{\mu}\phi$ and
$R^{\mu \nu}\partial_{\mu}\phi\partial_{\nu}\phi$ were used and the author found one de Sitter attractor solution; more recently, in Ref. \cite{Sushkov:2009hk}, it 
was found that the equation of motion for the scalar field can be reduced to a second order differential equation when it is kinetically
coupled to the Einstein tensor, $G^{\mu \nu}\partial_{\mu}\phi\partial_{\nu}\phi$; in Ref. \cite{Saridakis:2010mf}, where the author investigated the cosmological scenarios for this kind of coupling; and in reference \cite{Anisimov:2005ne} where a large class of Lagrangians of the form $L = Q(\Box\phi)$ was investigated, where $Q$ is a convex function. This lattermost theory allows for an inflationary evolution of the
universe driven  from rather generic initial conditions and for which, it has been called $B$-inflation or Box-inflation. 

Another earlier attempt to study the most general Higher Derivative scalar gravity is shown in  \cite{Elizalde:1994nz}. 
Yet another, in which one loop renormalization and asymptotic behavior of a higher derivative scalar theory in curved space-time can be seen in 
\cite{Elizalde:1994sn}. The conformal version of such theories is proposed therein, and asymptotic freedom is attempted as a solution to the ghost problem.

In this article, we would like to combine these ideas in a more simple setting for which the higher order term is calculated with a homogeneous FRW 
metric. This allows the transformation of a complex cosmological problem, where the lagrangian is of higher order in the time derivatives,  into a problem of classical mechanics. In order to do so, we use a coupling term inspired by the so-called Pais-Uhlenbeck (PU)
oscillator. This oscillator was proposed by Pais and Uhlenbeck as a non-localized action for solving the ultraviolet behavior of field theories \cite{Pais:1950za}. These kinds of theories are not free of problems, however; because the 
equations of motion are of the fourth-order,  there are ghosts therefore. 
These ghosts appear due to the linear instability (or Ostrogradsky linear instability) 
of the theory \cite{kat, Woodard:2006nt}. Concerning this instability, there is a no-go theorem, the so-called Ostrogradski theorem, which states that: if the higher order time derivatives Lagrangian is non-degenerate, there is at least one linear 
instability in the Hamiltonian \cite{ostro}.  The presence of ghosts usually spoils 
unitary and/or causality features of the theory, which is why higher derivatives theories are not usually considered good 
theories. To circumvent this problem, one may introduce an interaction term and show the existence of  a safe region in the parameters space where the theory is well behaved 
(as developed in Ref. \cite{Smilga:2004cy}). For examples of  PU oscillators in classical mechanics see, for instance \cite{Robert:2006nj,Smilga:2008pr,Smilga:2013vba}. 

Another possible way of dealing with the Ostrogradski ghost associated with non-degenerate higher order theories is based on an existing residual gauge symmetry that might be used to consistently select a stable physical Hilbert space \cite{Jimenez:2012ak}, interestingly such a field could be amplified during inflation and would give an effective cosmological constant today. This quantization procedure was motivated by previous works on gauge vector fields \cite{Jimenez:2008nm,Jimenez:2009dt} and the introduction of the associated St\"{u}ckelberg field. 
The first non-singular bounce model free of theoretical pathologies (such as ghosts, superluminality, graceful-exit issues, etc), was presented in \cite{Cai:2012va}. An interesting review about the topic of building a healthy bouncing/cyclic universe can be found in \cite{Cai:2014bea}.

\section{Smilga approach to classical mechanics}
First we would like to review the toy model  proposed by Smilga \cite{Smilga:2004cy} with equation of motion 
\begin{equation}
q^{IV}=\frac{d\alpha}{dt}, \label{eqm1}
\end{equation}
where $\alpha$ is some function of $q$, i.e, a potential. The above equation can be obtained from the higher-derivative action
\begin{equation}
S=\int dt\left(\frac{1}{2}\ddot{q}^2-\alpha(q)\right), \label{eqmaction}
\end{equation}
Since (\ref{eqm1})
 is of fourth order, the phase space is 4-dimensional. Therefore, we can describe the phase space with a pair of canonical variables and their momenta $(P_1,Q_1)$ and $(P_2,Q_2)$ with the Hamiltonian
\begin{equation}
H=P_1Q_1+\frac{P_2^2}{2}+\alpha(Q_1), \label{Ham}
\end{equation}
where one can always choose $\alpha(Q_1)$ to be some function which is bounded from below. The first term in (\ref{Ham}), which is linear in $P_1$, is the signal of the Ostrogradski linear instability. Since $P_1$ takes values throughout the phase space, there is no barrier preventing some degrees of freedom 
of the theory from having arbitrary negative energies. In other words, the Hamiltonian is not bounded from below. This corresponds 
to the Ostrogradski no-go theorem \cite{ostro}. Therefore, the higher order 
derivative Lagrangians always have  at least one linear instability, which leads to the presence of ghosts in the system. As said before, these  ghosts spoils the unitary and causality features of the theory so that these types of systems should therefore, at first glance, abandoned. Nevertheless, there is one kind of exorcism can try to do over the ghost. 

In this line of reasoning, then,  we would like to comment just two proposals. In Ref. \cite{Li:2005fm}, it was proved that the Ostrogradski instability can be removed by the addition of constraints, in which the original phase 
space of the theory is reduced.   On the other hand, Smilga \cite{Smilga:2004cy} found that a comparatively ``benign" mechanical 
higher-derivative system exists where the classical vacuum is stable under small perturbations and the problems appear 
only at non-perturbative levels. The author used the following example,
\begin{equation}
L=\frac{1}{2}\left((\ddot{q}+\Omega^2 q)^2-\frac{\alpha}{4}q^4-\frac{\beta}{2}q^2\dot{q}^2\right), \label{l1}
\end{equation} 
which corresponds to a higher-derivative model involving two kind of non-linear terms 
$\sim q^4$ and $\sim q^2\dot{q}^2$. This system is benign if the non-linear terms in the Lagragian have the opposite sign,  compared to the quadratic term $\sim (\ddot{q}+\Omega^2 q)^2$. As such it is expected that the system is benign if both $\alpha$ and $\beta$ are positive, and malicious if 
both $\alpha$ and $\beta$ are negative \cite{Smilga:2004cy}. This simplest example shows how the interaction (the coupling) term plays a decisive role in the benign or malicious behavior of the theory. 

In this paper, we propose a covariant model  with a minimal coupling between the scalar field and the geometry, but with higher-derivative terms inspired by the Pais-Uhlenbeck. 
The Pais-Uhlenbeck oscillator was proposed in \cite{Pais:1950za} for field
theories with non-localized action in order to correct the ultraviolet behavior of the theory. The action describing the PU oscillator is 
\begin{equation}
S=\frac{\gamma}{2} \int dt (\ddot{q}^2-(\omega_1^2+\omega_2^2)\dot{q}^2+\omega_1^2\omega_2^2 q^2),\label{PU}
\end{equation}
that leads to the equations of motion of fourth order
\begin{equation}
q^{IV}+(\omega_1^2+\omega_2^2)\ddot{q}^2+\omega_1^2\omega_2^2 q=0.\label{PUEM}
\end{equation}
Now, if we use the extra-coordinate $x=\dot{q}$ with the corresponding canonical momentum $P_x$, the canonical Hamiltonian is given by
\begin{equation}
H=P_qx+\frac{P_x^2}{2}+\frac{(\omega_1^2+\omega_2^2)x^2}{2}-
\frac{\omega_1^2\omega_2^2q^2}{2},\label{HPU}
\end{equation}
where the Ostrogradski instability encodes in the first term. The fourth order equation (\ref{PUEM}) gives a propagator like
\begin{equation}
G(E)=\frac{1}{(E^2+m_1^2)(E^2+m_2^2)}, \label{propa1}
\end{equation}
that can be rewritten as
\begin{equation}
G(E)=\frac{1}{m_2^2-m_1^2}\left(\frac{1}{E^2+m_1^2}-\frac{1}{E^2+m_2^2}\right). \label{propa2}
\end{equation}
Therefore, the PU oscillator is not free of the Ostrogradski instability and it exhibits ghost in its particle content. 

The next section is devoted to a cosmological construction based on the PU oscillator. We explore the malicious behavior of the ghost and its possible corrected by the interaction between geometry and the scalar field.

\section{Higher derivative coupling formulation}\label{s3}
First,  let us describe one simple model introduced in \cite{Li:2005fm} where the action of the system is given by
\begin{equation}
S=\int d^4x\sqrt{-g}\left(\frac{R}{2}-\frac{1}{2} g^{\mu \nu}\nabla_{\mu}\phi\nabla_{\nu}\phi+\frac{\alpha}{2}\Box \phi \Box \phi -V(\phi)\right),\label{action11}
\end{equation} that is,
a kind of Lee-Wick dark energy. In this case, the equation of motion for the scalar field is given by 
\begin{equation}
\Box \phi+ \alpha\, \Box^2 \phi-\frac{dV}{d\phi}=0,\label{eq11}
\end{equation}
and the corresponding energy momentum tensor is described by
\begin{align}
&T^{\mu \nu}=\left(\frac{1}{2}\nabla_{\rho}\phi \nabla^{\rho}\phi+ \frac{\alpha}{2}\,\Box \phi \Box \phi +\alpha\,\nabla^{\rho}\phi\nabla_{\rho}(\Box\phi)+V\right)g^{\mu \nu} \nonumber \\ & -\nabla^{\mu}\phi \nabla^{\nu}\phi- \alpha\,\nabla^{\nu}\phi\nabla^{\mu}(\Box\phi)-\alpha\, \nabla^{\mu}\phi\nabla^{\nu}(\Box\phi)
.\label{tmunu1}
\end{align}
Under the scalar field redefinition \cite{Li:2005fm,Creminelli:2005qk}
\begin{eqnarray}
\chi&=&\alpha\, \Box \phi, \\ \label{trans}
\psi&=&\phi+\chi,
\end{eqnarray}
the energy momentum tensor \eqref{tmunu1} can be written as
\begin{eqnarray}
T^{\mu \nu}&=& \frac{1}{2}\left(\nabla_{\rho}\psi\nabla^{\rho}\psi-\nabla_{\rho}\chi\nabla^{\rho}\chi+ 2V(\psi-\chi)+\frac{\chi^2}{\alpha}\right)g^{\mu \nu} \nonumber \\&-&
\nabla^{\mu}\psi\nabla^{\nu}\psi+\nabla^{\mu}\chi\nabla^{\nu}\chi
\end{eqnarray} 
and the corresponding Lagrangian is given by
\begin{equation}
\mathcal{L}=-\frac{1}{2}\nabla_{\mu}\psi\nabla^{\mu}\psi+\frac{1}{2}\nabla^{\mu}\chi\nabla^{\mu}\chi-V(\psi-\chi)-\frac{\chi^2}{2\alpha}.\label{la1}
\end{equation}
This means that the single-field higher derivative model is equivalent to a two field model where one field is conventional ($\chi$), and the second one is a ghost ($\psi$). This result is consistent with Ostrogradski's theorem. Therefore, the presence of ghosts in higher derivative cosmology is inevitable.

This class of models is closely related to the so-called quintom paradigm 
\cite{Feng:2004ad,Guo:2004fq,Zhang:2005eg,Aref'eva:2005fu,Zhao:2006mp,Lazkoz:2006pa,Vernov:2006dm,Lazkoz:2007mx,Setare:2008pz,Cai:2009zp,Aref'eva:2009xr,Leon:2012vt} through a Lee-Wick transformation of the kind seen in \eqref{trans}. It is important to mention that some cosmological features might be lost by the transformation since both fields are not independent. 

Our purpose is to investigate the lagrangian density corresponding to action \eqref{action11}, which is 
\begin{equation}
 \mathcal{L}=\frac{1}{2}\sqrt{-g}\left(R+\nabla_{\mu}\phi\nabla^{\mu}\phi+
\alpha\nabla_{\mu}\nabla^{\mu}\phi\nabla_{\nu}\nabla^{\nu}\phi-2 V(\phi)\right),
\end{equation}
where $\alpha$ is the coupling parameter. Additionally, we consider a radiation source with energy density $\rho_r=\rho_{r,0} a^{-4}$ as the background.  

For a homogeneous, isotropic, and spatially-flat 
universe, the line element is described by
\begin{equation}
 ds^2=dt^2-a(t)^2 d\mathbf{x^2}.
\end{equation}
Now we can use the fact that  Einstein's equations for an homogeneous, isotropic, and flat universe can be derived from a
pointlike Lagrangian \cite{Demianski:1991zv}:
\begin{equation}
L=L(a, \phi, \dot a, \dot\phi, \ddot a, \ddot\phi)=\frac{1}{2}\left[6 (a^2\ddot a+ a \dot a^2)+a^3 \dot\phi^2+\alpha a^3 \ddot\phi^2-2 a^3 V(\phi)-\frac{2\rho_{r,0}}{a}\right],
\end{equation} 
leading to the equations of motion
\begin{equation}
 \frac{\partial L}{\partial q^{i}}-\frac{d}{dt}\left(\frac{\partial L}{\partial \dot{q}^{i}}\right)+
\frac{d^2}{dt^2}\left(\frac{\partial L}{\partial \ddot{q}^{i}}\right)=0, \; \text{where} \; q^{i}=(a,\phi).   
\end{equation} 
So, the equations of motion for the scale factor and the  scalar field are respectively
\begin{equation}
 2a\ddot{a}+\dot{a}^2 +\frac{a^2}{2}\left(\dot{\phi}^2+\alpha\ddot{\phi}^2-2 V(\phi)\right)+\frac{\rho_{r,0}}{3 a^2}=0, \label{eins1}
\end{equation}
\begin{equation}
 \alpha\phi^{(IV)} +6\alpha\left(\frac{\dot{a}}{a}\right)^2\ddot{\phi}+3\alpha\left(\frac{\ddot{a}}{a}\ddot{\phi} +2\frac{\dot{a}}{a}\,\dddot{\phi}\right)-
3\frac{\dot{a}}{a}\dot{\phi}-\ddot{\phi}-\frac{\partial V}{\partial\phi}=0.\label{scalar}
\end{equation}

Since the lagrangian is not an explicit function of time, we can use the first Jacobi integral, or in other words, we can apply Noether's theorem for second order theories \cite{logan} with a lagrangian invariant under time translations, to get the conservation equation
\begin{equation}
-\rho_{0}=L-\dot{q}^{j}\frac{\partial L}{\partial \dot{q}^{j}}+\dot{q}^{j}\frac{d}{dt}\left(\frac{\partial L}{\partial\ddot{q}^{j}}\right)-
\ddot{q}^{j}\frac{\partial L}{\partial\ddot{q}^{j}}.
\end{equation}
Since our original system is covariant,  we can fix $\rho_0=0$ and obtain a Friedmann-like equation
\begin{equation}
 3\left(\frac{\dot{a}}{a}\right)^2=\frac{1}{2}\left(\dot{\phi}^2+\alpha\ddot{\phi}^2+2 V(\phi)\right)-
3\alpha\frac{\dot{a}}{a}\dot{\phi}\ddot{\phi}-\alpha\dot{\phi}\,\dddot{\phi}+\frac{\rho_{r,0}}{a^4}. \label{fried}
\end{equation}
Therefore the cosmological behavior of our system is described by the equation (\ref{eins1}), (\ref{scalar}) and the Friedmann-like
constraint (\ref{fried}). 

Additionally, we can define an effective Dark Energy (DE) source with energy density and pressure given by 
\begin{subequations}
\label{DE_sector}
\begin{align}
&\rho_{DE}:=\frac{1}{2} \dot\phi^2+\frac{1}{2} m^2 \phi^2+\frac{1}{2} \alpha  \ddot\phi^2-\frac{3 \alpha  \dot a \dot \phi \ddot \phi}{a}-\alpha \dddot \phi
   \dot\phi,\\
&p_{DE}:=	\frac{1}{2} \dot\phi^2-\frac{1}{2} m^2 \phi^2+\frac{1}{2} \alpha  \ddot\phi^2,
\end{align}
\end{subequations} where we have chosen a quadratic potential $V(\phi)=\frac{1}{2} m^2 \phi^2.$

Therefore, we can combine the Friedmann equations \eqref{fried} and
\eqref{eins1} in the usual form
 \begin{eqnarray}
\label{Fr1b}
3H^2& =& \rho_r + \rho_{DE}   \\
\label{Fr2b}
2\dot{H}& =&-\left(\frac{4}{3}\rho_r+\rho_{DE}+p_{DE}\right),
\end{eqnarray}
The conservation equation for radiation is 
\be\label{Cons1} \dot\rho_r=-4 H \rho_r. 
\ee

The dark energy density and pressure satisfy
the usual evolution equation
\begin{eqnarray}\label{Cons2}
\dot{\rho}_{DE} +3H(\rho_{DE}+p_{DE})=0,
\end{eqnarray}
and we can also define the dark energy equation-of-state parameter as
usual
\begin{eqnarray}\label{dde}
w_{DE}\equiv \frac{p_{DE}}{\rho_{DE}}.
\end{eqnarray}

Alternatively, we have defined the effective (total) equation of state parameter by 
\be\label{dde1}
w_{\text{eff}}\equiv \frac{p_{DE}+\frac{1}{3}\rho_r}{\rho_{DE}+\rho_r}.
\ee 

Finally, we introduce the dimensionless energy densities
\be\label{ode}
\Omega_{DE}\equiv \frac{\rho_{DE}}{3 H^2},
\ee

\be
\Omega_{r}\equiv \frac{\rho_r}{3 H^2},
\ee
which satisfy the Friedmann equation (\ref{Fr1b}).

In the next sections, we explore the parameter space to see the benign or malicious behavior of this system. 

\section{Qualitative behavior in the Phase space}

In this section, we perform stability analysis of the cosmological scenario at hand. In order to do that, we first transform it to its autonomous form \cite{Perko,Ellis,Copeland:1997et,Ferreira:1997au,Coley:2003mj,Chen:2008ft,Cotsakis:2013zha,Giambo':2009cc}
\be \label{eomscol}
\textbf{X}'=\textbf{f(X)}, \ee 
where $\bf{X}$ is a column vector of auxiliary variables,
and prime denotes derivatives
with respect to $N=\ln a$. From this, one extracts the  critical points
$\bf{X_c}$  which satisfy  $\bf{X}'=0$. In order to determine
their stability properties, one takes the Taylor expansion around them  up to first
order as \be\label{perturbation} \textbf{U}'={\bf{Q}}\cdot
\textbf{U},\ee with $\textbf{U}$, the
column vector of the perturbations of the variables and ${\bf {Q}}$, the matrix  containing the coefficients of the
perturbation equations. The eigenvalues of ${\bf {Q}}$ evaluated at the specific critical point determine their type and
stability.

\subsection{Phase space}\label{sect:4.1}
 
In our context the column vector denoted as $\textbf{X}$,  is given by
\begin{align} \label{aux}
x=\frac{\dot\phi}{\sqrt{6} H}, \; y=\frac{\ddot \phi}{\sqrt{6} H},\; z=\frac{m\phi}{\sqrt{6} H},\nonumber\\
u=\frac{H}{\dot\phi\ddot\phi},\; v=\frac{\alpha\dot\phi}{3  H^2}\left[\dddot\phi +3 H\ddot\phi\right], \Omega_r\equiv \frac{\rho_r}{3  H^2}, 
\end{align} which, with Friedman equation \eqref{fried}, are related through 
\be\label{variable_reduction}
v=-1+x^2+\alpha y^2 +z^2+\Omega_r.
\ee
Additionally, we introduce the new time variable $\tau=\ln a$, i.e., $f'\equiv \frac{d f}{d\tau}=\frac{\dot f}{H}$.
The evolution equations for \eqref{aux} are:
\begin{subequations}
\label{dyn_syst}
\begin{align}
\label{eqx}
&x'=\frac{3}{2} x \left(y^2 (\alpha +4 u)+x^2-z^2+1\right)+\frac{x \Omega_r}{2},\\
\label{eqy}
&y'=\frac{3}{2} y \left(2 u y^2+x^2-z^2-1\right)+\frac{3 u y \left(x^2+z^2-1\right)}{\alpha }+\frac{y \Omega_r (\alpha +6 u)}{2 \alpha }+\frac{3
	\alpha  y^3}{2},\\
\label{eqz} 
&z'=\frac{3}{2} \left(4 m u x^2 y+z \left(x^2+\alpha  y^2-z^2+1\right)\right)+\frac{z \Omega_r}{2},\\
\label{equ}
&u'=-\frac{3 u^2 \left(x^2+z^2-1\right)}{\alpha }-\frac{u \Omega_r (\alpha +6 u)}{2 \alpha }-\frac{3}{2} u \left(6 u y^2+x^2-z^2-1\right)-\frac{3}{2}
\alpha  u y^2,\\
\label{eqr}
&\Omega_r'=\Omega_r \left(3 x^2+3 \alpha  y^2+\Omega_r-3 z^2-1\right),\\
\label{eqv}
&v'=x^2 (6 y (2 m u z+3 u y+\alpha  y)+4 \Omega_r+3)+\nonumber \\ & +\left(\alpha  y^2+\Omega_r+z^2-1\right) \left(3 y^2 (\alpha +2 u)+\Omega_r-3
z^2\right)+3 x^4,
\end{align}
\end{subequations} where the prime denotes derivative with respect to $\tau$. 

The equation \eqref{variable_reduction} is preserved by the flow of \eqref{dyn_syst}, i.e., taking the time derivative on both sides, and using the evolution equations \eqref{dyn_syst} to get an identity. Thus, we can use the relation \eqref{variable_reduction} to eliminate one variable, $v$, whose evolution equation \eqref{eqv} is decoupled from the rest. From \eqref{eqx}, \eqref{eqy}, \eqref{equ} it follows that the signs of $x, y$ and $u$ are invariant. This means, e.g., that solutions with initial value $u(0)<0$ never cross the line $u=0$. Additionally, observe that the system is form invariant under the discrete symmetry $(x, y,   \Omega_r)\rightarrow (-x, -y, \Omega_r)$. However, it is not invariant under the changes $z\rightarrow - z$ and $u\rightarrow - u$. Finally, the fractional energy density $\Omega_r$ must be non-negative. With the above features combined, we can investigate the dynamics restricted to the reduced unbounded phase space $\Psi:=\{(x,y,z,u,\Omega_r)\in\mathbb{R}^5: x\geq 0, y\geq 0,   \Omega_r\geq 0\}$.

Now, in order to explain the physical meaning of the critical points of the autonomous system (\ref{dyn_syst}) we need to rewrite the cosmological parameters, defined in the previous section,  in terms of the dimensionless variables (\ref{aux}). Following this, the cosmological parameters (\ref{dde}), (\ref{dde1}) and (\ref{ode})  can be expressed as:
\begin{align}
&w_{DE}=\frac{x^2+\alpha  y^2-z^2}{1-\Omega_r},\\
&w_{eff}=x^2-z^2+\alpha y^2+\frac{\Omega_r}{3},\\
&\Omega_{DE}=1-\Omega_r,
\end{align}
while  the deceleration parameter becomes:\footnote{In order to avoid confusions, recall that in Section \ref{s3} we introduce $q^{i}$ as the set of generalised coordinates, while henceforth $q$, as usual,  represents the deceleration parameter.}
\begin{align}
&q=-\left[1+\frac{\dot{H}}{H^2} \right]=\frac{1}{2} \left(1+3 x^2+3 \alpha  y^2-3 z^2+\Omega_r\right).
\end{align}


In the following the dynamical behavior at the finite region are investigated. 
Then in table \ref{crit} the real and physically interesting
critical points of the autonomous system \eqref{dyn_syst} are presented. 

\begin{enumerate}
\item The curves of the singular points $P_1^\pm$ have effective cosmological parameters $w_{\text{eff}}=-1, q=-1$, i.e., each point on it behaves as de Sitter solutions. They are always saddle-like. First, it follows that $H=\frac{1}{6 x_c y_c u_c}\rightarrow \infty$ at the equilibrium point since $u_c=y_c=0$.  On the other hand, from the definitions of $z_c$ and $x_c$, it follows that  $\phi\sim H$ and $\dot\phi \sim H$, which  implies $\ddot\phi\sim u_c^{-1}\rightarrow \infty$ at equilibrium. Now, since $y$ goes to zero, it follows that $H$ must tend to infinity faster than $\ddot\phi$ does. 

\item The curves of the singular points $P_2^\pm$ have effective cosmological parameters $w_{\text{eff}}=-1, q=-1$, i.e., they behave as de Sitter solutions. They have a 3D stable manifold and a 2D center manifold. Henceforth, to investigate its stability we must resort to numerical experimentation or use sophisticated tools like  Center Manifold Theory. Since at equilibrium $x_c$ and $z_c$ are finite, it follows that $\dot\phi\sim H$ and $\phi\sim H$. Now, combining the definitions of $x$ and $u$, it follows that  $\ddot\phi= \frac{\sqrt{6}M}{6 x_c u_c},$ which, combined with $y_c=0$, implies that  $H$ must go to infinity as the equilibrium point is approached. 
  
\item $P_3$ is always a saddle critical point in the phase space. Its behaviour is independent  of whether the radiation is taken into account ($\Omega_r=0$). In this case, the effective DE component would mimic cold dark matter fluid ($w_{eff}=0$) at background level. 

\item $P_4$ mimics a stiff solution, i.e.,  $w_{\text{eff}}=1$. It is a source. All the derivatives of the scalar field, with the exception of $\ddot\phi$, go to infinity less quickly than $H$ does as time goes backward. 

\item $P_5$ is a radiation-dominated solution and is a saddle, as expected. 

\item $P_6$ mimics a matter-dominated solution with $w_{\text{eff}}=0$, i.e., it represents a dust solution, and is a saddle point.  At background level, it has the same behaviour as $P_3$.
\end{enumerate}

\begin{table*}
\begin{center}
{\begin{tabular}
{|c| c| c| c| }
 \hline
Cr. P./curve & $(x,y,z,u,\Omega_r)$ & $v$ &Existence \\
\hline\hline
$P_1^\pm$  & $\left(\sinh(\beta),0,\pm\cosh(\beta),0,0\right)$ & $2 \sinh^2(\beta)$ & always\\[0.2cm]
\hline
$P_2^\pm$  & $\left(\sinh(\beta),0,\pm\cosh(\beta),\frac{1}{2} \alpha \, \text{csch}^2(\beta),0\right)$ & $2 \sinh^2(\beta)$ & $\beta\neq 0$ \\[0.2cm]
\hline
$P_3$  & $\left(0,0,0,0,0\right)$ & $-1$ & always \\[0.2cm]
\hline
$P_4$  & $\left(0,\frac{\sqrt{\alpha}}{\alpha},0,0,0\right)$ & $0$ & $\alpha>0$ \\[0.2cm]
\hline
$P_5$  & $\left(0,0,0,0,1\right)$ & $0$ & always \\[0.2cm]
\hline
$P_6$  & $\left(0,0,0,-\frac{\alpha}{2},0\right)$ & $-1$ & always\\[0.2cm]
\hline
\hline
\end{tabular}}
\end{center}
\caption{\label{crit} The critical points of the autonomous system \eqref{dyn_syst}. }
\end{table*}

\begin{table*}
\begin{center}
{\begin{tabular}
{|c| c| c| c| c| c| c|}
 \hline
Cr. P./curve  & Eigenvalues &Stability  & $w_{DE}$ & $w_{\text{eff}}$  & q  & Cosmological solution \\
\hline\hline
$P_1^\pm$  & $-3,-3,-4,3,0$ & saddle  & $-1$ & $-1$ & $-1$ & de Sitter \\[0.2cm]
\hline
$P_2^\pm$   & $-3,-3,-4,0,0$ & nonhyperbolic  & $-1$ & $-1$ & $-1$ & de Sitter \\[0.2cm]
\hline
$P_3$  & $-1,-\frac{3}{2},\frac{3}{2},\frac{3}{2},\frac{3}{2}$ & saddle & $0$ & $0$ & $\frac{1}{2}$ & dust-like\\[0.2cm]
\hline
$P_4$  & $3,3,3,2,0$ & nonhyperbolic & $1$ & $1$ & $2$ & stiff-like \\[0.2cm]
\hline
$P_5$  & $-1, 1, 1, 2, 2$ & saddle & \_ & $\frac{1}{3}$ & $1$ & radiation-dominated \\[0.2cm]
\hline
$P_6$  &$-1,-\frac{3}{2},\frac{3}{2},\frac{3}{2},0$ & saddle & $0$ & $0$ & $\frac{1}{2}$ & dust-like\\[0.2cm]
\hline
\hline
\end{tabular}}
\end{center}
\caption{\label{crit1b} Stability conditions, cosmological parameters, and cosmological behavior of solutions for the critical points of the autonomous system \eqref{dyn_syst}.}
\end{table*}

\subsubsection{Evolution rates for the cosmological solutions near $P_2^\pm$}

For $P_2^\pm$, $u_c\neq 0, x_c\neq 0$. From the definitions of $u_c, v_c$, and $x_c$, the following relations are valid at the equilibrium point. 
\begin{subequations}
\begin{align}
&\frac{\dot\phi \ddot\phi}{H}=\frac{1}{u_c}\implies \frac{1}{H}\frac{d(\dot\phi^2)}{dt}=\frac{2 }{u_c}\implies \dot\phi^2= \ln \left(\frac{a}{a_0}\right)^{2/u_c},\\ 
&v_c=\frac{\alpha \dot\phi\ddot\phi}{3  H}\left[\frac{\dddot\phi}{H\ddot\phi}+3\right]=\frac{\alpha}{3 u_c}\left[\frac{\dddot\phi}{H\ddot\phi}+3\right]\implies \frac{d\ln \ddot\phi}{d\ln a}\equiv\frac{\dddot\phi}{H\ddot\phi}=3\left[\frac{u_c v_c}{\alpha}-1\right],\\
&\ddot\phi=\frac{\sqrt{6} }{6 x_c u_c},\\
&H=\frac{\sqrt{6}\dot\phi}{6   x_c}\implies H=\frac{\sqrt{6}\sqrt{\ln \left(\frac{a}{a_0}\right)^{\frac{2 }{u_c}}}}{6   x_c}.
\end{align}
\end{subequations}

Combining all the above expressions we obtain 
\begin{subequations}
	\label{eq4.11}
\begin{align}
&a(t)= a_0 \exp\left[\frac{1}{6\alpha}\left(\sqrt{6} a_1  \sinh(\beta) +t\right)^2\right],\\
&H(t)=\frac{\sqrt{6} a_1  \sinh(\beta)+t}{3 \alpha },\\
&\phi(t)=\frac{ t \sinh(\beta) \left(12 a_1
    \sinh (\beta )+\sqrt{6} t\right)}{6 \alpha },\\	
&\dot\phi(t)=\frac{ \sinh(\beta) \left(6 a_1  \sinh(\beta)+\sqrt{6} t\right)}{3 \alpha },\\
& \ddot\phi(t)= \frac{\sqrt{\frac{2}{3}} \sinh(\beta)}{\alpha },\\
&\dddot \phi(t)=0.
\end{align}
\end{subequations}

The energy density and pressure of DE at the equilibrium point given by 
{\small{
\begin{subequations}
\begin{align}
&\rho_{DE} = \frac{ \sinh ^2(\beta ) \left(8 \alpha +4 a_1 \sinh(\beta)  \left(6 a_1  \sinh(\beta)  \left(2 m^2 t^2-2\right)+\sqrt{6} t
   \left(2 m^2 t^2-4\right)\right)+2 m^2 t^4-8 t^2\right)}{24 \alpha ^2},\\
&p_{DE}= \frac{\sinh ^2(\beta ) \left(8 \alpha +4 a_1  \sinh(\beta) 
   \left(\sqrt{6} t \left(4-2 m^2 t^2\right)-6 a_1  \sinh(\beta)  \left(2 m^2 t^2-2\right)\right)-2 m^2 t^4+8 t^2\right)}{24 \alpha ^2}
\end{align}
\end{subequations}}}

In the case of $\alpha=0$,  the slow-roll quasi-de Sitter solution (which looks
similar to \eqref{eq4.11}), was first derived in \cite{Starobinsky1978}. Now, the relevant quantities associated with solution \eqref{eq4.11} are $\rho_{DE}, p_{DE}$ and $w_{DE}$, which, in the limit $\alpha<0, |\alpha|m^2\ll 1$, are given by:
\begin{subequations}
	\begin{align}
	&\rho_{DE}=-\frac{\sinh ^2(\beta ) \left(t^2-\alpha +2 a_1 \sinh (\beta ) \left(\sqrt{6} t+3 a_1 \sinh (\beta )\right)\right)}{3 \alpha
		^2}+\mathcal{O}\left(m^2 |\alpha| \right),\\
	& p_{DE}=\frac{\sinh ^2(\beta ) \left(t^2+\alpha +2 a_1 \sinh (\beta ) \left(\sqrt{6} t+3 a_1 \sinh
		(\beta )\right)\right)}{3 \alpha ^2}+\mathcal{O}\left(m^2 |\alpha| \right),\\
	& w_{DE}=-1+\frac{2 \alpha }{\alpha +2 a_1 \sinh (\beta ) \left(3 a_1 \sinh (\beta )+\sqrt{6} t\right)+t^2}+\mathcal{O}\left(m^2 |\alpha|\right).\label{4.13c}
	\end{align}
\end{subequations}
Now, concerning the duration of the metastable quasi-de Sitter stage, dynamical system techniques do not allow the exact duration of the lapse of time for the transition from one equilibrium point to the other to be obtained. However, from \eqref{4.13c}, it follows that the value of the effective dark energy is close to $-1$ for large enough values of $t$. A rough estimate of the duration of this phase can be inferred from investigating  the values of $t$ for which the equation of state parameter of Dark Energy remains in a small interval containing the value $w_{DE}=-1$. For example, given $\Delta>0$, we get 
$-1\leq w_{DE}<-1+\Delta$ for the choice $$a_1,\sinh (\beta )\in \mathbb{R},  \alpha <0, 0<\Delta <\frac{2 \alpha }{\alpha -6 a_1^2 \sinh ^2(\beta )}, t\geq
\sqrt{\frac{\alpha (\Delta -2) }{\Delta }}-\sqrt{6} a_1 \sinh (\beta ).$$ Although it is not a unique choice in leading to the same interval for $w_{DE}$. 

Now, let us examine the stability of $P_2^+$ using the center manifold theorem \cite{Perko}.
In order to prepare the system the analysis, we introduce the new variables 
\begin{subequations}
\begin{align}
&u_1= \frac{1}{8} \alpha  \coth (\beta ) \text{csch}(\beta ) \left(2 \alpha  m y (1-3 \cosh (2 \beta )) \text{csch}^2(\beta )-8 x
   \coth (\beta )+8 z\right),\\
&u_2= 12 \alpha ^2 m y \sinh ^2(\beta ) \cosh (\beta ),\\
&v_1= \frac{1}{8} \text{csch}^2(\beta )
   \left(2 \alpha ^2 m y (3 \cosh (2 \beta )-1) \coth (\beta ) \text{csch}(\beta ) +\right. \\ & \left. +4 (\alpha  \Omega_r+u \cosh (2 \beta )-u  -2 \alpha 
   \cosh (\beta ) (z-x \coth (\beta )))\right),\\
&v_2= 3 \alpha  \cosh ^2(\beta ) \left(\cosh (\beta ) \left(2 \alpha  m y-2 z\right)+2 x
   \sinh (\beta )-\Omega_r\right),\\
&v_3= \Omega_r,
\end{align}
\end{subequations}  which allows for the translation of $P_2^+$ to the origin $(u,v_1,v_2,v_3,v_4)=(0,0,0,0,0)$ and the system \eqref{dyn_syst} reduces to its Jordan real form. In this case, the Jordan form of the Jacobian matrix evaluated at the origin is
\be
\left(
\begin{array}{ccccc}
 0 & 1 & 0 & 0 & 0 \\
 0 & 0 & 0 & 0 & 0 \\
 0 & 0 & -3 & 1 & 0 \\
 0 & 0 & 0 & -3 & 0 \\
 0 & 0 & 0 & 0 & -4 \\
\end{array}
\right).
\ee 
Now, the center manifold of the origin is given locally by the graph
\be\left\{(u_1, u_2,v_1,v_2,v_3): v_i=h_i(u_1,u_2), h_i(0,0)=0, \mathbf{Dh}(\mathbf{0})=\mathbf{0}, i=1\ldots 4,  |(u_1,u_2)|<\delta\right\},\ee where $\delta$ is ``small'' and  $\mathbf{Dh}(\mathbf{0})$ denotes the matrix of derivatives evaluated at the origin. 
The functions $v_i$ must satisfy the set of quasilineal partial differential equations:
\begin{align}\label{quasilinear}
G_i(u_1,u_2, h_1, h_2, h_3) -\frac{\partial h_i}{\partial u_1}-\frac{\partial h_i}{\partial u_2}=0, \; i=1,2,3,
\end{align}
where $G_i(u_1,u_2, h_1, h_2, h_3)\equiv v_i'|_{v_i=h_i(u_1,u_2)}, \; i=1,2,3$, i.e., the expressions for evolution equations $v_i'$ after the replacement $v_i\rightarrow h_i(u_1,u_2).$  

Assuming that the functions $v_i$ can be expressed locally as 
\begin{subequations}\label{v_i's}
\begin{align}
&v_1=a_1 u_1^2 + a_2 u_1 u_2 + a_3 u_2^2 + \mathcal{O}(3),\\
&v_2=b_1 u_1^2 + b_2 u_1 u_2 + b_3 u_2^2 + \mathcal{O}(3),\\
&v_3=c_1 u_1^2 + c_2 u_1 u_2 + c_3 u_2^2 + \mathcal{O}(3),
\end{align}
\end{subequations} where $\mathcal{O}(3)$ denotes terms of 3rd order, it is possible to solve the system \eqref{quasilinear} up to third order. 
Substituting the expressions \eqref{v_i's} in  \eqref{quasilinear}  and comparing the coefficients of the same powers of $u_1$ and $u_2$, we obtain the relations for the $a_i$'s, $b_i$'s and $c_i$'s:
\begin{subequations}
\label{coefs}
\begin{align}
&a_1= \frac{11 \sinh ^2(\beta )+4 \tanh ^2(\beta )-1}{8 \alpha }-6 a_2 \sinh ^4(\beta )+\frac{b_2}{2},\\ 
&a_3= \frac{(-3
   \cosh (2 \beta )+28 \cosh (4 \beta )-5 \cosh (6 \beta )+28) \text{csch}^{10}(\beta ) \text{sech}^2(\beta )}{18432 \alpha }+\nonumber \\ 
	& -\frac{1}{144}
   \text{csch}^4(\beta ) \left(12 a_2+b_2 \text{csch}^4(\beta )\right)+\frac{\cosh (2 \beta ) \text{csch}^8(\beta ) \text{sech}^2(\beta )}{288
   \alpha ^2 m^2},\\
&b_1= \frac{3 \sinh ^4(\beta ) (1-3 \cosh (2 \beta ))}{4 \alpha }-6 b_2 \sinh ^4(\beta ),\\
&b_3= \frac{\left(7
   \text{csch}^4(\beta )+6 \text{csch}^2(\beta )-5\right) \text{csch}^2(\beta )}{192 \alpha }-\frac{1}{12} b_2 \text{csch}^4(\beta
   )+\frac{\text{csch}^4(\beta )}{48 \alpha ^2 m^2},\\
&c_1= -8 c_2 \sinh ^4(\beta ),\\
&c_3= -\frac{1}{16} c_2 \text{csch}^4(\beta
   )
\end{align}
\end{subequations}
Thus, the graph of the center manifold of the origin is given by the functions \eqref{v_i's} with the coefficients given by \eqref{coefs}.
 
\begin{figure}[t]
	\centering
		\includegraphics[width=0.5\textwidth]{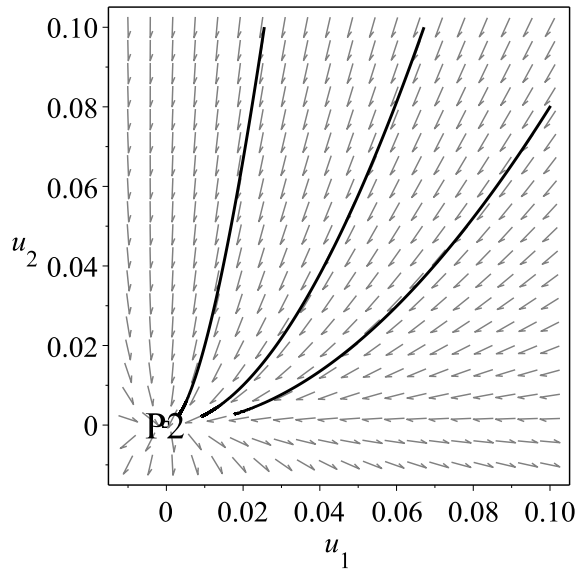}
	\caption{Phase space of the system \eqref{center_evol} for $\alpha=1, m=\frac{\sqrt{2}}{2}, \beta=1$. The line $u_2=0$ is invariant for the flow. The orbits above the line, corresponding to the portion of the phase space $u\geq 0$, are attracted by the origin. The orbits below this line depart from the origin.}
	\label{fig:Fig1}
\end{figure}

\begin{figure}[t]
	\centering
		\includegraphics[width=0.5\textwidth]{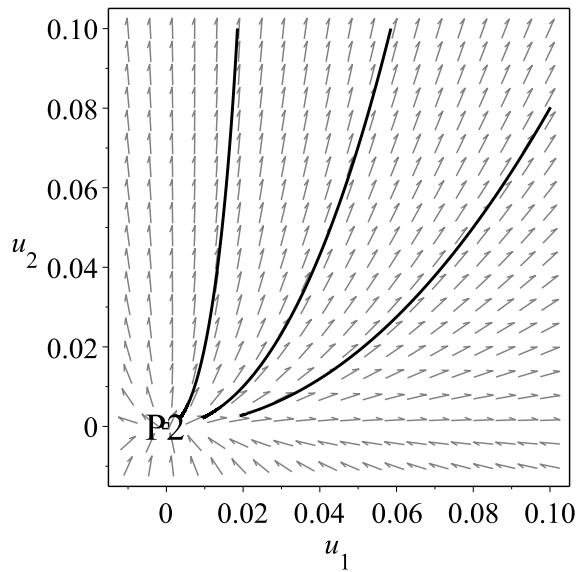}
	\caption{Phase space of the system \eqref{center_evol} for $\alpha=-1, m=\frac{\sqrt{2}}{2}, \beta=1$. The line $u_2=0$ is invariant for the flow. The orbits above the line, corresponding to the portion of the phase space with $u\geq 0$, depart from the origin. The orbits below this line are attracted by the origin.}
	\label{fig:Fig2}
\end{figure}

Plugging \eqref{coefs} back  into the evolution equations for $u_1$ and $u_2$, we obtain that the evolution on the center manifold is given by 
\begin{subequations}
\label{center_evol}
\begin{align}
&u_1'=\frac{u_2 \text{csch}^2(\beta ) \left(-48 \alpha  m^2 u_1-u_2 \text{csch}^6(\beta ) \left(2\alpha m^2+4\right)+2\alpha  m^2
   u_2 \text{csch}^4(\beta )\right)}{192 \alpha ^2 m^2}+ \mathcal{O}(3),\\
&u_2'=	-\frac{u_2^2 \text{csch}^2(\beta )}{2 \alpha }+ \mathcal{O}(3),
\end{align}
\end{subequations} 
Neglecting the 3rd order terms, we obtain the general solution \eqref{center_evol}:
\begin{subequations}
\begin{align} 
&u_1(\tau)=\frac{c_2}{\sqrt{-\alpha  c_1 \cosh (2 \beta )+\alpha  c_1+\tau }}+\frac{\text{csch}^4(\beta ) \left(-2\alpha m^2 \cosh (2 \beta )+6 \alpha 
   m^2+8\right)}{48 m^2 \left(-\alpha  c_1 \cosh (2 \beta )+\alpha  c_1+\tau \right)},\\
&u_2(\tau)=\frac{2 \alpha }{\tau  \text{csch}^2(\beta )-2 \alpha  c_1}.
\end{align}
\end{subequations} 
The equations \eqref{center_evol} define a local flow, i.e., a flow defined for all $\tau \geq 2 \alpha  c_1 \sinh ^2(\beta )$ but not for the whole real line. For $\alpha>0$, it is easy to prove that for $u_2(t_0)>0$ the origin is approached when $\tau\rightarrow +\infty.$ Solutions with $u_2 < 0$ depart from the origin. In figure \ref{fig:Fig1} the typical behavior of solutions on the center manifold of $P_2^+$ is displayed.  For the numerics, we choose $\alpha=1, m=\frac{\sqrt{2}}{2}, \beta=1$. For $\alpha<0$, the typical behavior is the time reverse of the above (see figure \ref{fig:Fig2}). 

For analyzing $P_2^-$ we introduce the new variables
{\small{
\begin{subequations}
\begin{align}
& u_1= \frac{1}{4} \alpha  \coth (\beta ) \text{csch}(\beta ) \left(2 \alpha  m y \left(\text{csch}^2(\beta )+3\right)-4 x \coth
   (\beta )-4 z\right),\\
&u_2= -12 \alpha ^2 m y \sinh ^2(\beta ) \cosh (\beta ),\\
&v_1= \frac{1}{4} \left(\alpha  \text{csch}(\beta )
   \left(2 \Omega_r \text{csch}(\beta )+\coth (\beta ) \left(-2 \alpha  m y \left(\text{csch}^2(\beta )+3\right)+4 x \coth (\beta )+4
   z\right)\right)+4 u\right),\\
&v_2= -3 \alpha  \cosh ^2(\beta ) \left(\cosh (\beta ) \left(2 \alpha  m y-2 z\right)-2 x \sinh (\beta
   )+\Omega_r\right),\\
&v_3= \Omega_r.
	\end{align}
\end{subequations}}}
Applying the center manifold theorem analogously as before, we obtain as a result that the dynamics on the center manifold are governed by 
the same system \eqref{center_evol}. Thus, the results proceed from the previous analysis. That is, for $\frac{H(t_0)}{\dot\phi(t_0)\ddot\phi(t_0)}>0$, $P_2^-$ is the attractor solution. 

\subsubsection{Evolution rates for the cosmological solutions near $P_4$}

For $P_4$ we have $y_c= \frac{\sqrt{\alpha}}{\alpha}.$ This point exists only for $\alpha>0$. From the definition of $y$, it follows that
\begin{equation}\label{P5a}
\frac{d\dot\phi}{d\ln a}\equiv \sqrt{6}  y_c \implies \dot \phi=\ln\left[\left(\frac{a}{a_0}\right)^{\sqrt{6}  y_c}\right].
\end{equation}
Taking successive time derivatives of the above expression gets 
\begin{subequations}\label{P5b}
\begin{align}
&\ddot\phi=\sqrt{6}  y_c \left(\frac{\dot a}{a}\right)^2,\\
&\dddot\phi=-\sqrt{6}  y_c \left(\frac{\dot a}{a}\right)^2+\sqrt{6}  y_c \frac{\ddot a}{a}.
\end{align}
\end{subequations}
Using the definition $H=\frac{\dot a}{a},$ and substituting the expressions \eqref{P5a} and \eqref{P5b} back  into the definition of $v$, we obtain at the equilibrium point 
\begin{equation}\label{P5c}
\frac{\ln\left[\left(\frac{a}{a_0}\right)^{\sqrt{6}  y_c}\right]\left(2\dot a ^2 +a \ddot a \right)}{ \dot a}=0.
\end{equation}
Solving the differential equation \eqref{P5c} we obtain the solution 
\begin{subequations}
\begin{align}
& a(t)=a_1 (t-t_0)^{\frac{1}{3}},\\
& H(t)=\frac{1}{3(t-t_0)},\\
&\phi(t)=-\frac{1}{3} (t-t_0) \left(\sqrt{6}  y_c-3 \ln \left[\left(\frac{a_1 \sqrt[3]{t-t_0}}{a_0}\right)^{\sqrt{6}
   y_c}\right)\right],\\
&\dot\phi(t)= \ln\left[\left(\frac{a_1}{a_0}(t-t_0)^{\frac{1}{3}}\right)^{\sqrt{6}  y_c}\right],\\
&\ddot \phi(t) =\frac{\sqrt{\frac{2}{3}}  y_c}{t-t_0},\\
&\dddot \phi(t) =-\frac{\sqrt{\frac{2}{3}}  y_c}{(t-t_0)^2}.
\end{align}
\end{subequations} where $y_c=\frac{\sqrt{\alpha}}{\alpha}.$

For this point, the energy density and pressure of the DE is given by 
\begin{equation}
\rho_{DE}=p_{DE}=\frac{1}{3 (t-t_0)^2}+ \ln\left[\left(\frac{a_1 \sqrt[3]{t-t_0}}{a_0}\right)^{\sqrt{6} 
   y_c}\right]+\mathcal{O}\left((t-t_0)^2\right).
\end{equation} That is, a stiff solution. 

This solution, corresponding to a big-bang singularity, is closely related to the general solution obtained in \cite{Leon:2008de} in the context of nonminimally coupled scalar field dark energy models. 

Now, let us examine the stability of $P_4$ using the center manifold theorem \cite{Perko}.
The center manifold of $P_4$ is tangent to the center subspace, the $u$-axis. 
Defining the new variables
\be u=u, v_1=\Omega_r, v_2=z, v_3=y+\frac{\sqrt{\alpha}}{2\alpha}\left(\Omega_r-2\right), v_4=x, \ee it is possible to translate $P_4$ to the origin $(u,v_1,v_2,v_3,v_4)=(0,0,0,0,0)$ and the system \eqref{dyn_syst} reduces to its Jordan real form. 
The center manifold of the origin is now given locally by the graph
\be\left\{(u,v_1,v_2,v_3,v_4): v_i=h_i(u), h_i(0)=0, h_i'(0)=0, i=1\ldots 4,  |u|<\delta\right\},\ee where $\delta$ is ``small''. 
The functions $h_i$ can be locally expressed as
$v_i=\alpha_{i 1} u^2+\alpha_{i 2} u^3 + \ldots \alpha_{i n} u^n +\mathcal{O}(u^{n+1}).$
Using the center manifold theorem we obtain that the graph is \be\left\{(u,v_1,v_2,v_3,v_4): v_i=\mathcal{O}(u^{n+1}), h_i(0)=0, \mathbf{Dh}(0)=\mathbf{0}, i=1\ldots 4,  |u|<\delta\right\},\ee where $\delta$ is ``small'', and the evolution equation on the center manifold is 
\be \label{centerP4}
u'=-\frac{6 u^2}{\alpha} + \mathcal{O}(u^{n+1}).
\ee 
The equation \eqref{centerP4} is a gradient-like equation with potential $U(u)=\frac{2 u^3}{\alpha}.$ From our previous analysis we know that the sign of $u$ is invariant. Thus, for $\alpha>0$, the solutions starting with $u(0)>0$ approach the origin as time goes forward. The solutions starting with $u(0)<0$ depart asymptotically from the origin. Thus, if we restrict our attention to the halfspace $u>0,$ point $P_4$ behaves like a saddle point (the center manifold attracts an open set of orbits). However, considering the evolution in the whole space, the origin is unstable and $P_4$ is a local source.

\subsection{Two-field model reformulation}

In order to express the model as a 2-field theory we introduce the scalar field redefinition:
\begin{equation}
\psi=\phi+\alpha\Box\phi,\; \chi=\alpha\Box\phi.
\end{equation}
Then, the system \eqref{Fr1b}, \eqref{Fr2b}, \eqref{Cons1} and \eqref{Cons2}, reduces to 
\begin{subequations}
\begin{align}
&\ddot \chi = -3 H \dot\chi+\frac{\chi}{\alpha }-m^2(\psi-\chi),\\
&\ddot \psi= -3 H  \dot\psi-m^2(\psi-\chi),\\
&\dot H=-\frac{1}{2} \left(\dot\psi^2-\dot\chi^2\right)-\frac{2}{3}\rho_r,\\
&\dot \rho_r=-4 H \rho_r,\\
\label{Fried2}
& 3H^2=\frac{1}{2} {\dot\psi}^2-\frac{1}{2} {\dot\chi}^2+\frac{\chi^2}{2 \alpha }+\frac{m^2}{2}(\psi-\chi)^2+\rho_r.
\end{align}
\end{subequations}
which is equivalent to a quintom field ($\psi$ quintessence and $\chi$ phantom) with potential 
\begin{equation}
U(\psi,\chi)=\frac{\chi^2}{2 \alpha }+\frac{m^2}{2}(\psi-\chi)^2
\end{equation}  with a radiation field included. 
The DE energy density and pressure are now written as
\begin{subequations}
\begin{align}
&\rho_{DE}:=\frac{1}{2} \left[m^2 (\chi-\psi)^2+\frac{  \chi^2}{\alpha }- {\dot\chi}^2+ {\dot \psi}^2\right],\\
&p_{DE}:=\frac{1}{2} \left[-m^2 (\chi-\psi )^2-\frac{  \chi^2}{\alpha }- {\dot\chi}^2+  {\dot\psi}^2\right]
\end{align}
\end{subequations}

It is well-known that under the field redefinition \cite{Li:2005fm}
\begin{equation}
	\phi_1=\frac{a_2 \chi-a_1 \psi}{a_1^2-a_2^2},
	\phi_2=\frac{a_1 \chi- a_2 \psi}{a_1^2-a_2^2}
\end{equation}
where 
\begin{equation}
	a_1=\frac{\sqrt{4 \alpha  m^2+1}-1}{2 \sqrt[4]{4 \alpha  m^2+1}},
	a_2=\frac{\sqrt{4 \alpha  m^2+1}+1}{2 \sqrt[4]{4 \alpha  m^2+1}},
	\end{equation}
we obtain two independent modes $\phi_1$ and $\phi_2$ that evolve independently in the universe, i.e.,
\begin{subequations}
	\begin{align}
	&\ddot{\phi}_1 +3 H \dot{\phi_1}-m_1^2 \phi_1=0,\\
	&\ddot{\phi}_2 +3 H \dot{\phi_2}+m_2^2\phi_2=0,
	\end{align}
\end{subequations}
 where we have defined the effective masses for fields $\phi_1$ and $\phi_2$, respectively,
 \begin{equation}
 	m_1^2=\frac{\sqrt{4 \alpha  m^2+1}+1}{2 \alpha },
 	m_2^2=\frac{\sqrt{4 \alpha  m^2+1}-1}{2 \alpha }.
  \end{equation}
The energy density of dark energy is rewritten as 
\begin{equation}
\rho_{DE}=-\frac{1}{2}\dot{\phi_1}^2+\frac{1}{2}\dot{\phi_2}^2+\frac{1}{2} m_1^2 \phi_1^2+\frac{1}{2}m_2^2\phi_2^2,
\end{equation}  
in other words, $\phi_1$ is a phantom mode\footnote{In the limit $\alpha  m^2\ll 1$ we obtain $a_1\approx 0, a_2\approx 1$, $m_1^2\approx \frac{1}{\alpha}, m_2^2\approx 0$ and $\chi\approx -\phi_1, \psi\approx \phi_2$ and the motion equations are completely integrable in a flat spacetime.}. 

As was shown in \cite{Li:2005fm}, there are no unphysical instabilities at the classical level associated with perturbations in  $\phi_1$. The spatial fluctuations with wavenumber $k>m_1$ are stable\footnote{The solution is oscillatory in time \cite{Li:2005fm}.}, with the exception of large scales $L>m_1^{-1}$ where a time-rising behavior takes place. For $m_1<H$ there are no instabilities inside the horizon \cite{Hsu:2004vr,Buniy:2005vh}.  The rising behaviors of the super-horizon modes of the phantom are supressed since Hubble expansion provides a friction force preventing these modes from increasing exponentially and as a result the instability is benign \cite{Smilga:2004cy}. 

It can also be proved that, for interval $\alpha<0, 4|\alpha|m^2\leq 1$, the solutions of the wave equation for $\phi$ do not grow exponentially in a flat space time. Furthermore, in the region $\alpha<0, |\alpha|m^2 \ll 1$, the ghost mass is $m_1^2\approx \frac{1}{|\alpha|}\gg m^2$, namely, the mass of the ghost exceeds the mass of the scalar field $\phi\equiv\frac{\phi_1+\phi_2}{\sqrt[4]{4 \alpha  m^2+1}}\approx \phi_1+\phi_2$ and the mass of the normal scalar particle $\phi_2$ is $m_2^2\approx 0$. Thus, it may be argued that the ghost  is benign, in the sense that it is difficult to excite it at the classical level.

\subsubsection{Phase space}

Let's introduce the normalized variables
\begin{equation}\label{VARS2}
\Omega_r=\frac{\rho_{r}}{3 H^2},u_1=\frac{\dot\chi}{\sqrt{6} H},u_2=\frac{\dot\psi}{\sqrt{6}
   H},u_3=\frac{m \chi}{\sqrt{6} H},u_4=\frac{m\psi}{\sqrt{6}H},u_5=\frac{\sqrt{2} m}{H},
\end{equation}
which are related through 
\begin{equation}\label{NewFR1}
\mu u_3^2-u_1^2+u_2^2+ (u_3-u_4)^2+\Omega_r=1.
\end{equation} where we have introduced the new parameter $\mu=\frac{1}{\alpha m^2}.$

The new variables \eqref{VARS2} are related to the old ones \eqref{aux}
by the non-linear transformation of coordinates
\begin{subequations}
\label{DirectTransf}
\begin{align}
&x=u_2-u_1,\\
&y=\frac{3 \sqrt{2} m (u_1-u_2)}{u_5}-\mu  m u_3,\\
&z=u_4-u_3,\\
&u=-\frac{u_5^2}{6 m^2
	(u_1-u_2) \left(6 u_1-6 u_2-\sqrt{2} \mu  u_3 u_5\right)},\\
&v=2 (u_1-u_2) \left(\frac{12 u_3^2
	(u_2-u_1)}{u_5^2}+u_1\right)-\frac{12 (u_1-u_2)^2 \left(u_1^2-u_2^2+2
	\left((u_3-u_4)^2-1\right)\right)}{\mu  u_5^2}
\end{align}
\end{subequations}
with inverse transformation
\begin{subequations}
\label{InverseTransf}
\begin{align}
&u_1=\frac{2 u y^2 \left(-6 \mu ^2 m^4 u^3 v y^2+\mu  m^2 u \left(x^2-2 z^2+2\right)-2 u y^2-2\right)-1}{4 \mu  m^2 u^2 x y^2 \left(6
	\mu  m^2 u^2 y^2-1\right)},\\
&u_2= \frac{2 u y^2 \left(-6 \mu ^2 m^4 u^3 y^2 \left(v-2 x^2\right)-\mu  m^2 u \left(x^2+2 z^2-2\right)-2 u
	y^2-2\right)-1}{4 \mu  m^2 u^2 x y^2 \left(6 \mu  m^2 u^2 y^2-1\right)},\\
&u_3=-\frac{2 u y^2+1}{2 \mu  m u y},\\
&u_4=z-\frac{2 u y^2+1}{2
	\mu  m u y},\\
&u_5=6 \sqrt{2} m u x y.
\end{align}
\end{subequations}
The variables \eqref{VARS2} are suitable for describing a portion of the solution space than cannot be accessed by the set of coordinates \eqref{aux}. The transformations \eqref{DirectTransf} (resp. \eqref{InverseTransf}) are not smooth for $u_5=0$ (resp. $u=0, x=0, y=0$), and so are not smooth at the fixed points. Thus, the critical points obtained for the coordinate system \eqref{VARS2} are indeed new points. 
Additionally, the new set of variables \eqref{VARS2} is more suitable for the numerics than \eqref{aux}, since for the variables \eqref{aux}, the variable $u$ and the variables $x,y$ take numerical values with several orders of magnitude of difference. Thus, it is worth investigating the solution space described by \eqref{VARS2}.

The evolution equations for \eqref{VARS2} are
\begin{subequations}
\label{dyn_syst2}
\begin{align}
&u_1'=-u_1^3+u_1 \left(u_2^2-2 \mu  u_3^2-2 (u_3-u_4)^2-1\right)+\frac{u_5 (\mu 
	u_3+u_3-u_4)}{\sqrt{2}},\\
&u_2'=-u_2 \left(u_1^2+2 \left(\mu 
u_3^2+(u_3-u_4)^2\right)+1\right)+u_2^3+\frac{u_5 (u_3-u_4)}{\sqrt{2}},\\
&u_3'=u_3
\left(-u_1^2+u_2^2-2 u_4^2+2\right)+\frac{u_1 u_5}{\sqrt{2}}-2 (\mu +1) u_3^3+4 u_3^2 u_4,\\
&u_4'=u_4
\left(-u_1^2+u_2^2-2 \left((\mu +1) u_3^2-1\right)\right)+\frac{u_2 u_5}{\sqrt{2}}+4 u_3 u_4^2-2
u_4^3,\\
&u_5'=-u_5 \left(u_1^2-u_2^2+2 \left(\mu  u_3^2+(u_3-u_4)^2-1\right)\right). 
\end{align}
\end{subequations}
where we have used the equation \eqref{NewFR1} as a definition of $\Omega_r.$

The equations \eqref{dyn_syst2} define a flow on the unbounded phase space 
\begin{equation}
\left\{(u_1,u_2,u_3,u_4,u_5)\in\mathbb{R}^5:0\leq \mu u_3^2-u_1^2+u_2^2+ (u_3-u_4)^2\leq 1\right\}.
\end{equation}

Finally, the cosmological parameters read 
\begin{subequations}\label{OBS2}
\begin{align}
&\Omega_{DE}=-u_1^2+u_2^2+\mu  u_3^2+(u_3-u_4)^2,\\
&\omega_{DE}=\frac{u_1^2-u_2^2+\mu  u_3^2+(u_3-u_4)^2}{u_1^2-u_2^2-\mu  u_3^2-(u_3-u_4)^2},\\
&\omega_{eff}=	\frac{1}{3} \left(-2 u_1^2+2 u_2^2-4 \mu 
   u_3^2-4 (u_3-u_4)^2+1\right),\\
&q=-u_1^2+u_2^2-2 \mu  u_3^2-2 (u_3-u_4)^2+1
\end{align}	
	\end{subequations}
	
	Table \ref{crit_2}  presents the critical points of the autonomous system \eqref{dyn_syst2}, and
	table \ref{crit_2b} presents the stability conditions, cosmological parameters, and cosmological behavior of solutions for them. 

\begin{table*}
\begin{center}
{\begin{tabular}
{|c| c| c| c| }
 \hline
Cr. P./curve & $(u_1,u_2,u_3,u_4,u_5)$ & $\Omega_r$ &Existence \\
\hline\hline
$Q_1$  & $\left(\sinh(\beta), \cosh(\beta), 0, 0, 0\right)$ & $0$ & always\\[0.2cm]
\hline
$Q_2^\pm$  & $\left(0, \pm 1, 0, 0, 0\right)$ & $0$ & always\\[0.2cm]
\hline
$Q_3^\pm$  & $\left(0, 0, u_{3c}, u_{3c}\pm\sqrt{1- \mu u_{3c}^2}, 0\right)$ & $0$ & $\mu\leq 0$ or \\
& & & $\mu> 0, -\sqrt{\frac{1}{\mu}}\leq u_{3c} \leq \sqrt{\frac{1}{\mu}}$ \\
[0.2cm]
\hline
$Q_4^\pm$  & $\left(0, 0, 0, \pm 1, 0\right)$ & $0$ & always\\[0.2cm]
\hline
$Q_5$  & $\left(0, 0, 0, 0, 0\right)$ & $1$ & always\\[0.2cm]
\hline
\hline
\end{tabular}}
\end{center}
\caption{\label{crit_2} The critical points of the autonomous system \eqref{dyn_syst2}. }
\end{table*}

\begin{table*}
\begin{center}
{\begin{tabular}
{|c| c|  c|  c| c| c|c |}
 \hline
Cr. P./curve  & Eigenvalues &Stability  & $w_{DE}$ & $w_{\text{eff}}$  & q & Cosmological solution  \\
\hline\hline
$Q_1$  & $3,3,3,2,0$ & unstable & $1$ & $1$ & $2$ & stiff-like   \\[0.2cm]
\hline
$Q_2^\pm$   & $3, 3, 3, 2, 0$  & unstable  & $1$ & $1$ & $2$ & stiff-like  \\[0.2cm]
\hline
$Q_3^\pm$  & $-4, -3, -3, 0, 0$ & nonhyperbolic & $-1$ & $-1$ & $-1$ & de Sitter \\[0.2cm]
\hline
$Q_4^\pm$  & $-4, -3, -3, 0, 0$ & nonhyperbolic & $-1$ & $-1$ & $-1$ & de Sitter \\[0.2cm]
\hline
$Q_5$  & $2, 2, 2, -1, -1$ &  saddle & \_ & $\frac{1}{3}$ & $1$ & radiation-dominated \\[0.2cm]
\hline
\hline
\end{tabular}}
\end{center}
\caption{\label{crit_2b} Stability conditions, cosmological parameters, and cosmological behavior of solutions for the critical points of the autonomous system \eqref{dyn_syst2}. }
\end{table*}

\begin{figure}
\centering
\includegraphics[width=0.65\textwidth, angle=90]{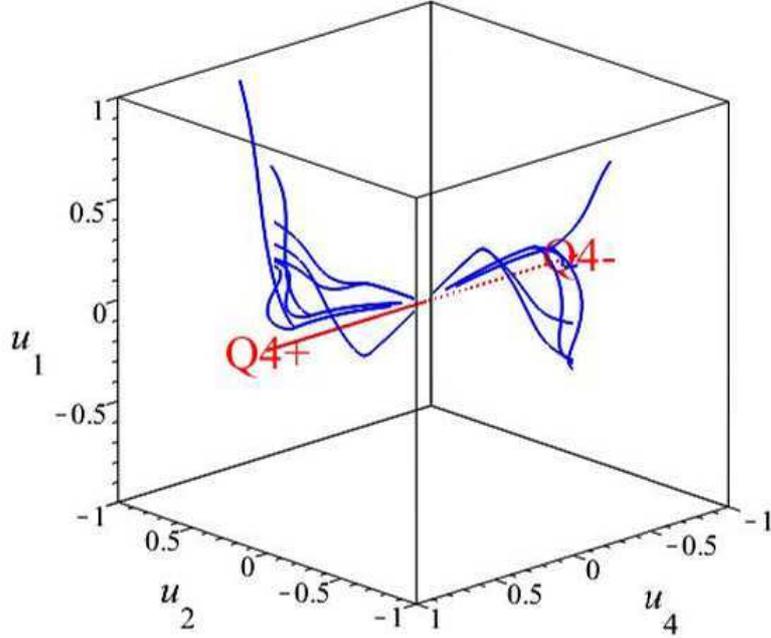}
\caption{Some projections of orbits of the phase space \eqref{dyn_syst2} for the choice $\alpha=0.1, \sqrt{2} m=0.1$. The horizontal solid (red) line corresponds to $Q_3^+$ and the horizontal dotted (red) line corresponds to $Q_3^-$. Both lines, representing de Sitter solutions, attracts an open set of orbits of \eqref{dyn_syst2}. The figure suggest that $Q_4^\pm$ are saddles.}
\label{fig:Fig7}
\end{figure}

Let us enumerate the critical points and critical curves of the system \eqref{dyn_syst2}:
\begin{enumerate}
\item $Q_1$ is a curve of points corresponding to stiff matter which are unstable. They correspond to the past attractor of the system \eqref{dyn_syst2}.
\item The critical points $Q_2^\pm$ belong to curve $Q_1$, and thus have the same dynamical behavior and the same physical interpretation of the whole curve of critical points.  
\item Points $Q_3^\pm$ exist for $\mu\leq 0$ or $\mu> 0, -\sqrt{\frac{1}{\mu}}\leq u_{3c} \leq \sqrt{\frac{1}{\mu}}$, and have effective cosmological parameters $w_{eff} =
-1, q = -1$, i.e., they behave as de Sitter solutions. They have a 3D stable manifold
and a 2D center manifold. Henceforth, to investigate its stability we must resort
to numerical experimentation or use sophisticated tools like the Center Manifold
Theory. Figure \ref{fig:Fig7} presents some projections of orbits of the phase space \eqref{dyn_syst2} for the choice of parameters  $\alpha=0.1, \sqrt{2} m=0.1$. The horizontal solid (red) line corresponds to $Q_3^+$ and the horizontal dotted (red) line corresponds to $Q_3^-$. Both lines, representing de Sitter solutions, attract an open set of orbits of \eqref{dyn_syst2}.

\item Points $Q_4^\pm$ always exist, and are special points of the curve $Q_3^\pm$.
 The effective cosmological parameters are $w_{eff} =
-1, q = -1$, i.e., they behave as de Sitter solutions. The simulation presented in figure \ref{fig:Fig7} suggests that they are saddles. More accurate characterization require the use of the Center Manifold Theory.

\item Point $Q_5$ always exists and corresponds to a radiation-dominated solution. As expected, it has saddle behavior, so it cannot attract the universe at late time, but rather corresponds to a transient epoch in cosmic history. 
\end{enumerate}

Finally, introducing the Poincar\'e variables:
	\begin{align}
\left\{U_1, U_2, U_3, U_4, U_5\right\}	&=\frac{1}{\sqrt{1+u_1^2+u_2^2+u_3^2+u_4^2+u_5^2}} \left\{u_1, u_2, u_3, u_4, u_5\right\},\end{align} 
  
we find that critical points of the system \eqref{dyn_syst2} at the infinite region are contained on the Poincar\'e hypersphere $S:= \{U_1^2+U_2^2+U_3^2+U_4^2=1,U_5=0\}$ and all the critical points on the finite region satisfy $u_5=0$. That is, all the possible stationary behavior of our model occurs on the regime $H\gg m$. Now, from the points located on the hypersphere $S$, the physical ones, that is, those inside the region  \begin{align*}
& R:=\left\{(U_1,U_2, U_3, U_4, U_5): -U_1^2+U_2^2+\mu  U_3^2+(U_3-U_4)^2\geq 0, \right. \nonumber \\ &
\left.  2 U_2^2+(\mu +2) U_3^2-2 U_3 U_4+2 U_4^2+U_5^2\leq 1,  \right. \nonumber \\ &
\left. U_1^2+U_2^2+U_3^2+U_4^2+U_5^2\leq 1\right\},
\end{align*} must  satisfy $2 U_1^2+U_3 (2 U_4-\mu  U_3)=1$. The complete stability analysis of the points at infinity is outside the scope of the present research.    

\section{Crossing the phantom divide}

The crossing of the phantom divide, i.e., that the equation of state parameter of DE crosses the value $w_{DE}=-1$, is possible for both $\alpha>0$ and $\alpha<0.$ Additionally, cyclic behavior appears for $\alpha<0$. In this section, we present some numerics for illustrating our analytical results. 

\subsection{Case $\alpha>0$}

In this section we present some numerical solutions and the regimes that appear for the case $\alpha>0$. 

\begin{figure}[h]                                                   
  \centering
    \includegraphics[width=.9\textwidth]{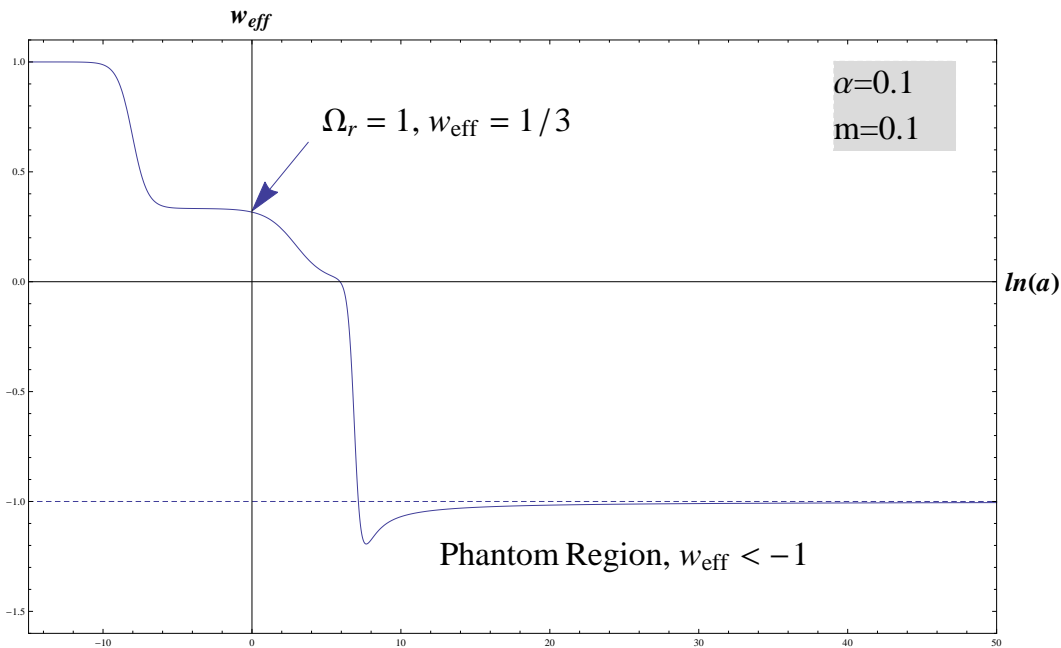}
  \caption{\label{Fig6} Evolution of $\omega(\tau)$, where $\tau=\ln a$, for $\alpha=0.1, \sqrt{2} m=0.1$. We set $\rho_r>0$.}
  \label{fig6}             
\end{figure}

Observe in figure \ref{fig6}, that the crossing of the phantom divide occurs once, and that the equation of state parameter keeps below this line all the time, before reaching asymptotically towards the de Sitter solution from below. This result is qualitatively the same for every $m$ and $\alpha$, both positive.  

\subsection{Case $\alpha<0$}

In this section, we discuss  the crossing of the phantom barrier $w_{DE}=-1$, and the cyclic behavior that appears for $\alpha<0$ for three different regimes  $|\alpha| \sim m$, $|\alpha| \gg m$ and $|\alpha| \ll m$.

\subsubsection{Numerical Solutions and Regimes}
\begin{figure}[h]
  \centering
    \includegraphics[width=.9\textwidth]{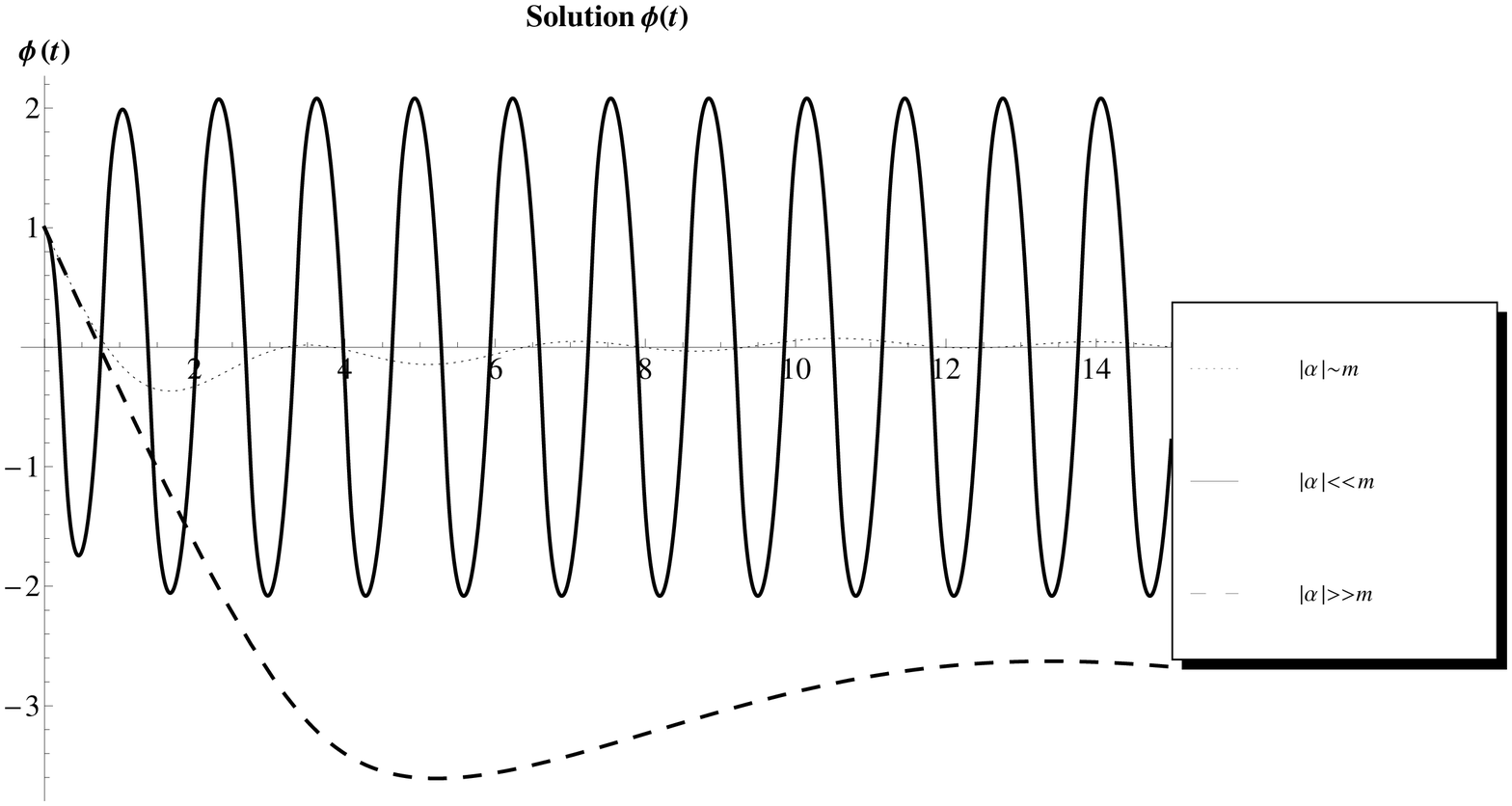}
  \caption{Evolution of $\phi(t)$ in the three different regimes  $|\alpha| \sim m$, $|\alpha| \gg m$ and $|\alpha| \ll m$. We choose values where $\alpha<0.$ We set $\rho_r=0$.}
  \label{fig_phi}
\end{figure}

\begin{figure}[h]
  \centering
    \includegraphics[width=.9\textwidth]{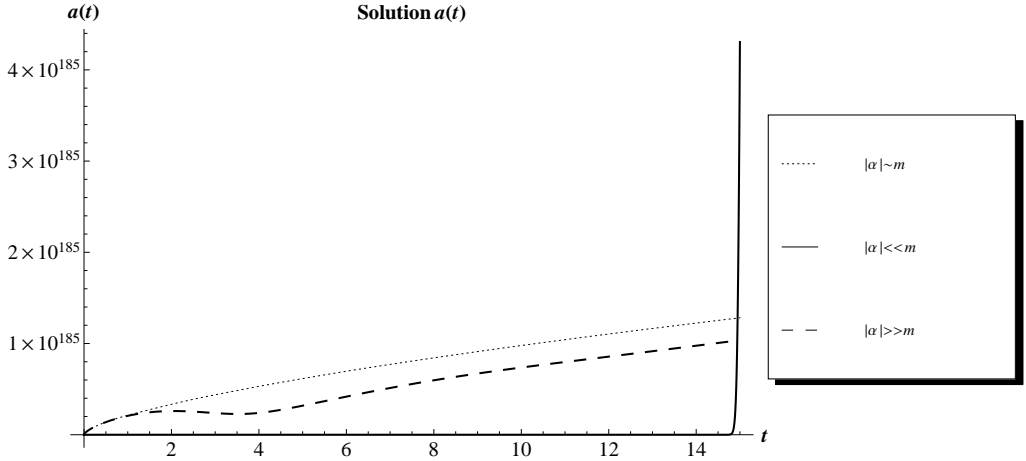}
  \caption{Evolution of $a(t)$ in the three different regimes  $|\alpha| \sim m$, $|\alpha| \gg m$ and $|\alpha| \ll m$. We choose values where $\alpha<0.$ The solutions for $|\alpha| \sim m$ and $|\alpha| \gg m$ are magnified by a factor of $5\times 10^{185}$ to be displayed in the same diagram. We set $\rho_r=0$.}
  \label{fig_a}
\end{figure}

\begin{figure}[h]                                                   
  \centering
    \includegraphics[width=.9\textwidth]{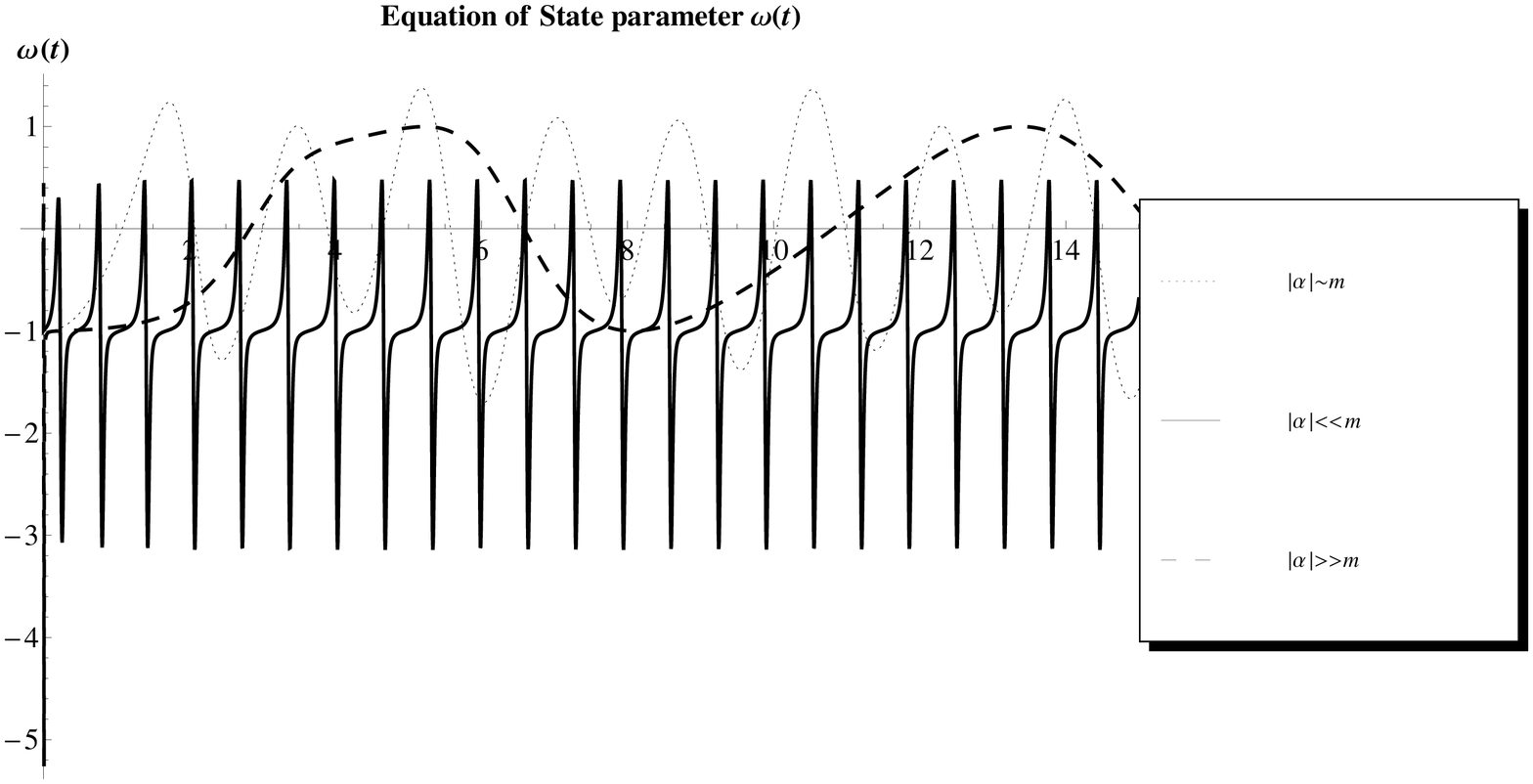}
  \caption{\label{Fig4} Evolution of $\omega(t)$ in the three different regimes  $|\alpha| \sim m$, $|\alpha| \gg m$ and $|\alpha| \ll m$. We choose values where $\alpha<0.$ We set $\rho_r=0$.}
  \label{fig}
\end{figure}

It is known that higher derivative terms involve ghosts \cite{Smilga:2004cy,Cai:2008qw}, 
but in some regimes of the theory the  ghosts are benign \cite{Smilga:2004cy}, that is,  they lead to a metastable vacuum. For illustration,
we plotted the numerical solutions when $V(\phi)=\frac{1}{2}m^2\phi^2$ in three different regimes: first, when $|\alpha|$ (the parameter
associated with the quartic derivative of $\phi$) is approximately equal to the parameter (mass) associated with the self interaction term,
 $m$; when $|\alpha| \gg m$; and finally, $|\alpha| \ll m$. The numerical solutions for the scalar field and the scale factor are drawn in figures
\ref{fig_phi} and \ref{fig_a}, respectively. Finally, figure \ref{Fig4} presents the evolution of $\omega(t)$ in the three different regimes $|\alpha| \sim m$, $|\alpha| \gg m$, and $|\alpha| \ll m$. We choose values where $\alpha<0.$ 


\section{Final Remarks}
We have considered a four-dimensional cosmology theory where the
scalar field is minimally coupled to gravity along with a
self-interacting  potential and includes a higher derivative term of the scalar field. 
Using the dynamical systems approach, we have obtained that for $\alpha>0$, and for initial values $$\frac{H(t_0)}{\dot\phi(t_0)\ddot\phi(t_0)}>0,$$ the system is attracted by 
the curve of singular points $P_2^\pm$ and corresponds to de Sitter solutions ($w_{\text{eff}}=-1, q=-1$). Since, at equilibrium, $x_c$ and $z_c$ are finite, it follows that $\dot\phi\sim H$ and $\phi\sim H$. Now, combining the definitions of $x$ and $u$, it follows that  $\ddot\phi$ is finite, which, combined with $y_c=0$, implies that  $H$ must go to infinity as the equilibrium point is approached. 
Additionally, the past attractor is very likely to be stiff solution, with $$\rho_{DE}=p_{DE}=\frac{1}{3 (t-t_0)^2}+ \ln\left[\left(\frac{a_1 \sqrt[3]{t-t_0}}{a_0}\right)^{\sqrt{6} 
   \frac{\sqrt{\alpha}}{\alpha}}\right]+\mathcal{O}\left((t-t_0)^2\right),$$ which represents a Big-bang singularity and is closely related to the general cosmological solution obtained in the context of nonminimally coupled scalar field dark energy
models.

For completeness, we have explored the relation of our model with a 2-field theory introducing scalar field redefinition. We introduced a set new coordinates suitable for describing a portion of the solution space that cannot
be accessed by the original coordinates. The stability of the de Sitter solutions is also studied. 

For $\alpha>0$, the crossing of the phantom divide occurs once, and
while the equation of state parameter keeps below this line, before asymptotically reaching towards
 the de Sitter solution from below. 

Now, for $\alpha<0$, we have found that the interaction allows  benign behavior in the scalar field, where the vacuum is metastable to be obtained. Namely, for $|\alpha|  \sim m$ the solutions of the equations of
motion shown the scalar field oscillating and being damped through the time period i.e., where ghosts are benign. For this regime, we see the
 scale factor solution accelerate  as usual. For $|\alpha|  \ll m$, we see an oscillating scalar field where the amplitude is not
damped during the regime, the scale factor does not accelerate for a period of time, and then accelerates abruptly. Finally for the case of $|\alpha|  \gg m$, the
scalar field  oscillates with a period longer than that of the time in which it displays the properties of benign ghosts. The scale factor accelerates, decelerates and then accelerates again after a short time. For  $\omega(t)$ in these three different regimes, we have the behaviors shown in figure \ref{Fig4}. For $|\alpha|  \sim m$ and $|\alpha|  \ll  m$, the phantom divide is crossed periodically, and for $|\alpha|  \gg m$, the phantom divide is crossed once, but as, the equation of state becomes greater than $-1,$ possible future crossings are less often. 

It is worth mentioning that, although we have just considered radiation and the Pais-Uhlenbeck modification here, we have obtained two solutions, $P_3$ and $P_6$, where the higher derivatives modification mimics a dust fluid, and both of them are saddle points. Thus, these solutions are candidates for the transient matter dominated era that preceded the current accelerated expansion phase. However, a more complete scenario should include both a radiation field and a dust fluid and is given by the pointlike lagrangian density 
\begin{equation} L=L(a, \phi, \dot a, \dot\phi, \ddot a, \ddot\phi)=\frac{1}{2}\left[6 (a^2\ddot a+ a \dot a^2)+a^3 \dot\phi^2+\alpha a^3 \ddot\phi^2-2 a^3 V(\phi)-\frac{2\rho_{r,0}}{a}-2 \rho_{m,0}\right], 
\end{equation}
where $\rho_{r,0}$ and $\rho_{m,0}$ are constants, and indices $r$ and $m$ mean radiation and matter, respectively. The investigation of this extension of our model is the aim of a subsequent paper, in which we will examine if the extended model allows for a {\em complete cosmological dynamic}, i.e., the existence of a viable radiation dominated era (RDE), a matter dominated era (MDE) and then a late time accelererated era \cite{Avelino:2013wea,Leon:2012vt}. At each of these stages, some form of matter dominates the dynamics, and is translated into different critical points which are connect by heteroclinic orbits, starting at a source and ending at a sink or attractor (see Refs. \cite{Ellis,Coley:2003mj,Heinzle:2007kv,UrenaLopez:2011ur}, for recent discussions on the role of heteroclinic orbits in Cosmology).  The heteroclinic orbits corresponding to a specific cosmological history where a RDE precedes a MDE and allows for a late time accelerated expansion are the targets of this analysis. Additionally, one has to study the perturbations and test it with astrophysical data.

\section*{Acknowledgments}

This paper is dedicated to the memory of our colleague and great friend, Sergio del Campo, first Chilean theoretical cosmologist, who sadly passed away. 

We would also like to thank Miguel Cruz, Nathalie Deruelle, Justo Lopez, Efrain Rojas, Adolfo Toloza, and Ricardo Troncoso for valuable discussions. Thanks are due to  Jose Beltran Jimenez, Yi-Fu Cai, Sergei Odintsov, Emmanuel N.~Saridakis, and Andrei V. Smilga for bringing our attention to useful references. This work was funded by Comisi\'{o}n Nacional de Investigaci\'{o}n Cient\'{i}fica y Tecnol\'{o}gica (CONICYT)
 through: FONDECYT Grant 1110076 (J.S.), DI-PUCV Grant 123713 (J.S.), FONDECYT Grant 3140244 (G.L.), DI-PUCV Grant 123730 (G.L.) and by FONDECYT Grant 11140309 (Y.L.). Y.L. thanks the PUCV for supporting
him through Proyecto DI Postdoctorado 2014. One of us (J.S.) wishes to thank the Department of Physics, Shanghai Jiao Tong University, and Prof. Bin Wang in particular  for his kind hospitality.
The authors thanks referees and editors whose comments helped to improve the original manuscript.

\end{document}